# A Variation of Dirac Equation, Based on SO(2,1) Group, with Applications to Low Dimensional Systems

Y.Ben-Aryeh

Physics Department Technion-Israel Institute of Technology, Haifa, 32000, Israel

e-mail: phr65yb@ph.technion.ac.il

**Abstract** A variation of Dirac equation based on SO(2,1) group is suggested for treating low dimensional systems in the three dimensional x,y,t space. Non-unitary representations are developed in an analogous way to those used in the ordinary Dirac equations and quantum field theory is developed for the present SO(2,1) relativistic equation. The theory is applied for low dimensional systems including especially holes-electrons pairs, or other two-particle states, in the low dimensional space.

## 1.Introduction

Leinaas and Myrheim [1-2] raised the idea that the two possibilities of identical particles to be either bosons or fermions, corresponding respectively to symmetric and antisymmetric wavefunctions is valid only for cases in which the particles move in a three dimensional spatial x,y,z space. For identical particles moving in the plane, the identical particles may follow a different statistics referred to as anyons [3]. The idea is that a particle circulating around an identical one in the three-dimensional spatial space can be continuously deformed into the identity, so that the wavefunction after the circulation should be equal to the original one. A full circulation is equivalent to two successive particle exchanges so that a single exchange leads to a phase factor either $1$ for bosons, or $-1$ for fermions. When we restrict the movement of the particles to two spatial dimensions there are more possibilities [4]. When a particle circulates around identical one in the plane it is not possible to deform its movement continuously to the identity, as intersection with the other particle is not topologically allowed. (For treating topological effects in quantum optics (see, e.g. [5]). In such case the wavefunction may acquire a phase which is different from $0$ or $\pi$. Such phases lead in the general case of many particles to a statistics given by the braid group (see, e.g [4]).

In the present work we study the problem of relativistic physics in the plane by using quantum field theory in which the ordinary Dirac equation, based on SO(3,1) group is reduced to a relativistic equation based on SO(2,1) group. The special result which appears in the present formalism is that the wavefunctions of this relativistic equation are two-dimensional in comparison with the wavefunctions of the ordinary Dirac equation which are four dimensional. The quantization of the SO(2,1) relativistic equation in the $x, y, t$ coordinates is made with quantum field methods analogous to those used for the ordinary Dirac equation [6-9] in the $x, y, z, t$ coordinates.

By using non-relativistic approximation for the present SO(2,1) Dirac equation, we get results which are consistent with the methods used for quantum Hall effects [10-12], in the plane. In the relativistic case we show that although the one particle operators satisfy the anti-commutation relations, two-particle states as hole-electron pairs, or other forms of pairs, can be described as bosonic systems [13].

## 2. The Wavefunctions for the SO(2,1) Relativistic Equation

We assume that the SO(2,1) relativistic equation which is a function of the $x, y, t$ coordinates can be given as:

$$\left[-i\frac{\sigma_0}{c}\frac{\partial}{\partial t} + \sigma_1\frac{\partial}{\partial x} + \sigma_2\frac{\partial}{\partial y} + \frac{mc}{\hbar}I\right]\Psi_\pm = 0 \quad . \tag{1}$$

Here $\sigma_1, \sigma_2$ and $\sigma_0 \equiv \sigma_3$ are the Pauli matrices, $m$ the electron mass, $c$ the light velocity, $I$ the unit $2\times 2$ matrix and $\Psi_\pm$ are the two-dimensional wavefunctions solutions of Eq. (1). This equation is analogous to the Dirac equation in the four $x, y, z, t$ coordinates. Using the definitions

$$\gamma_0 = -i\sigma_0 \quad , \quad \gamma_1 = \sigma_1 \quad , \quad \gamma_2 = \sigma_2 \quad , \quad x_0 = ct \quad , \quad x_1 = x \quad , \quad x_2 = y \quad . \tag{2}$$

(1) can be written in analogy to the ordinary Dirac equation as

$$\left(\gamma_\mu \frac{\partial}{\partial x_\mu} + \frac{mc}{\hbar}I\right)\Psi_\pm = 0 \quad , \quad \mu = 0, 1, 2 \quad . \tag{3}$$

The mathematical reasons for assuming such equation are as follows:

a) $\sigma_1, \sigma_2$ and $\sigma_3$ satisfy the Clifford algebra which has been essential for deriving the ordinary Dirac equation:

$$\{\sigma_i, \sigma_j\} = \delta_{i,j} \quad , \quad (i, j = 1, 2, 3) \quad , \tag{4}$$

where the curled brackets denote the usual anti-commutation relation, assuming that Pauli two-dimensional matrices replace the four dimensional Dirac $\gamma$ matrices.

b) We define:

$$K_1 = \frac{-i\sigma_2}{2} \quad , \quad K_2 = \frac{i\sigma_1}{2} \quad , \quad K_3 = \frac{\sigma_0}{2} \quad , \tag{5}$$

and then we get the commutation-relations of the SO(2,1) algebra :

$$[K_1, K_2] = -iK_3 \quad , \quad [K_3, K_1] = iK_2 \quad , \quad [K_2, K_3] = iK_1 \quad . \tag{6}$$

Using the definition of Eq. (5) the SO(2,1) relativistic equation can be written as

$$\left[ \frac{-2iK_3}{c} \frac{\partial}{\partial t} - 2iK_2 \frac{\partial}{\partial x} + 2iK_1 \frac{\partial}{\partial y} + \frac{mc}{\hbar} I \right] \Psi = 0 \quad . \tag{7}$$

(1), (3) and (7) express the same relativistic equation in different forms. (1) will be used in the present Section for deriving the wavefunctions solutions of the SO(2,1) relativistic equation and their ortho-normalization relations. The quantization of this relativistic equation, according to quantum field theory, will be developed in Section 3. In Section 4 we discuss certain possible applications of the present theory.

c) The unitary representations of the Lorentz group have been discussed in various works (see, e.g. [14]). We should notice that there are not any finite dimensional unitary representations of the Lorentz group and the Lorentz group solutions are finite only as non-unitary representations. Thus we are forced to take non-unitary representations which for the $SO(2,1)$ group representations can be taken as two-dimensional. We assume that the SO(3,1) Lorentz group can be reduced to the SO(2,1) Lorentz group when we restrict the coordinates to those of the $x, y, t$ space and it leads to certain new non-unitary representations. While there is an extensive literature on the unitary representations of the SO(2,1) group (see. e.g. [15]) the use of non-unitary representations, in relation to (1), has not been exploited.

d) By squaring the operators in (1), in front of the $\Psi_\pm$ fields we get:

$$-\frac{1}{c^2} \frac{\partial^2}{\partial t^2} \Psi = -\left( \frac{\partial^2}{\partial x^2} + \frac{\partial^2}{\partial y^2} \right) \Psi + \left( \frac{mc}{\hbar} \right)^2 \Psi \quad , \tag{8}$$

which is the wave equation of the Klein-Gordon form.

e) The solutions of (1) can be given as :

$$\Psi_+ = u(k_x, k_y) \exp\left[ i(k_x x + k_y y - \omega t) \right] \quad , \tag{9}$$

$$\Psi_- = v(k_x, k_y) \exp[-i(k_x x + k_y y - \omega t)] \quad , \tag{10}$$

where $u$ and $v$ are two-dimensional wavefunctions which are functions of $k_x$ and $k_y$, where $k_x$ and $k_y$ are the components of the wavevector in the $x$ and the $y$ direction, respectively, $\omega$ is the frequency, and the homogeneous equations obtained by substituting (9-10) into (1) have solutions only under the condition

$$\omega^2 = (k_x^2 + k_y^2)c^2 + \left(\frac{mc^2}{\hbar}\right)^2 \quad . \tag{11}$$

(11) can be transformed into the form

$$E^2 = (p_x^2 + p_y^2)c^2 + (mc^2)^2 \quad , \quad (E = \hbar\omega \quad , \quad p_x = \hbar k_x \quad , \quad p_y = \hbar k_y), \tag{12}$$

so that the relativistic equation (11) or equivalently (12) is obtained.

Let us evaluate in the following analysis the classical solutions of (1), by following the above theoretical scheme. In the rest frame $(k_x = k_y = 0)$ we get

$$\left[-i\frac{\sigma_0}{c}\frac{\partial}{\partial t}\right]\Psi_\pm + \left(\frac{mc}{\hbar}I\right)\Psi_\pm = 0 \quad , \tag{13}$$

and the solutions are

$$\begin{aligned}
\Psi_+ &= \exp\left[-i\left(\frac{mc^2}{\hbar}\right)t\right]\begin{pmatrix}1\\0\end{pmatrix} \Rightarrow \hbar\omega = mc^2 \quad , \\
\Psi_- &= \exp\left[i\left(\frac{mc^2}{\hbar}\right)t\right]\begin{pmatrix}0\\1\end{pmatrix} \Rightarrow \hbar\omega = -mc^2
\end{aligned} \tag{14}$$

We encounter already here the problem of obtaining a negative energy solution for the state $\psi_-$, often referred to as 'hole', but such holes will be related later in Section 3 to positrons by using the quantum field theory.

One should notice that (9-10) include the generalization of (14) to the moving system, which includes boosts in the $x$ and $y$ directions, and where the frequency $mc^2/\hbar$ is generalized to the frequency given by (11). For the non-relativistic case we can use the approximation $(k_x^2 + k_y^2)c^2 \ll (mc^2/\hbar)^2$, but the negative energy solutions cannot be neglected in solid state physics, where such negative energy solutions represent 'holes' (or in another way positrons).

For evaluating $\Psi_+$, in the general case $(k_x \neq 0, k_y \neq 0)$, we assume that the *un-normalized* solution of $u(k_x, k_y)$ can be given as

$$u(k_x, k_y) = \begin{pmatrix} 1 \\ G_1(k_x, k_y) \end{pmatrix} \quad , \tag{15}$$

so that $\Psi_+$ of (1) is given as

$$\Psi_+ = \begin{pmatrix} 1 \\ G_1(k_x, k_y) \end{pmatrix} \exp\left[i(k_x x + k_y y - \omega t)\right] \quad . \tag{16}$$

For evaluating $G_1(k_x, k_y)$ we use (16) and the relations:

$$-i \frac{\sigma_0}{c} \frac{\partial}{\partial t} \Psi_+ = \begin{pmatrix} -\omega/c & 0 \\ 0 & \omega/c \end{pmatrix} \Psi_+ \quad , \tag{17}$$

$$\left(\sigma_1 \frac{\partial}{\partial x} + \sigma_2 \frac{\partial}{\partial y}\right) \Psi_+ = \begin{pmatrix} 0 & i(k_x - ik_y) \\ i(k_x + ik_y) & 0 \end{pmatrix} \Psi_+ \quad . \tag{18}$$

By substituting (16-18) into (1), and multiplying by $c$, we get:

$$\begin{bmatrix} -(\omega - mc^2/\hbar) & ic(k_x - ik_y) \\ ic(k_x + ik_y) & (\omega + mc^2/\hbar) \end{bmatrix} \begin{pmatrix} 1 \\ G_1 \end{pmatrix} = 0 \quad . \tag{19}$$

The homogeneous equations (19) have solutions only under the condition that the determinant of the coefficients in the square brackets vanishes, which leads to (11)

Separating (19) into its components we get:

$$ic(k_x - ik_y) G_1 = (\omega - mc^2/\hbar) \quad , \tag{20}$$

$$-ic(k_x + ik_y) = (\omega + mc^2/\hbar) G_1 \quad . \tag{21}$$

From (21) we get

$$G_1 = \frac{-i(k_x + ik_y)c}{(\omega + mc^2/\hbar)} = \frac{-i(p_x + ip_y)c}{E + mc^2} \quad . \tag{22}$$

By substituting (22) into (20) we verify the validity of this equation.

For evaluating $\Psi_-$, in the general case, we assume that the *un-normalized* solution of $v(k_x, k_y)$ can be given as

$$v(k_x, k_y) = \begin{pmatrix} G_2(k_x, k_y) \\ 1 \end{pmatrix} \quad , \tag{23}$$

so that $\Psi_-$ of (1) is given as

$$\Psi_- = \begin{pmatrix} G_2(k_x, k_y) \\ 1 \end{pmatrix} \exp\left[-i(k_x x + k_y y - \omega t)\right] \quad . \tag{24}$$

Here, again, $k_x$ and $k_y$ are the components of the wavevector in the $x$ and $y$ directions, $\omega$ is the frequency, and $G_2$ is a function of $k_x, k_y$.

For evaluating $v(k_x, k_y)$ we use the relations:

$$-i\frac{\sigma_0}{c}\frac{\partial}{\partial t}\Psi_- = \begin{pmatrix} \omega/c & 0 \\ 0 & -\omega/c \end{pmatrix}\Psi_- \quad , \tag{25}$$

$$\left(\sigma_1\frac{\partial}{\partial x} + \sigma_2\frac{\partial}{\partial y}\right)\Psi_- = \begin{pmatrix} 0 & -i(k_x - ik_y) \\ -i(k_x + ik_y) & 0 \end{pmatrix}\Psi_- \quad . \tag{26}$$

By substituting Eq. (24-26) for $\Psi_-$ into (1), and multiplying with $c$, we get:

$$\begin{pmatrix} (\omega + mc^2/\hbar) & -i(k_x - ik_y)c \\ -i(k_x + ik_y)c & -(\omega - mc^2/\hbar) \end{pmatrix}\begin{pmatrix} G_2 \\ 1 \end{pmatrix} = 0 \quad . \tag{27}$$

Here, again, the homogeneous equations (27) have solutions only under the condition that the determinant of the coefficients in the square brackets vanishes, which leads again to (11).

Separating (27) into its components we get:

$$ic(k_x - ik_y) = (\omega + mc^2/\hbar)G_2 \quad , \tag{28}$$

$$-ic(k_x + ik_y)G_2 = (\omega - mc^2/\hbar) \quad . \tag{29}$$

From (28) we get:

$$G_2 = \frac{i(k_x - ik_y)c}{(\omega + mc^2/\hbar)} = \frac{i(p_x - ip_y)c}{(E + mc^2)} \tag{30}$$

By substituting (30) into (29), we verify the validity of this equation.

One should notice that the above wavefunctions are based on non-unitary representations so that they satisfy special ortho-normalization relations given as follows. The ortho-normalization relations are obtained between the two-dimensional vectors $u(k_x, k_y)$ and $v(k_x, k_y)$, and $u(k_x, k_y)^\dagger$ and $v(k_x,k_y)^\dagger$, where $u(k_x,k_y)^\dagger = \bar{u}(k_x,k_y)\sigma_0$ and $v(k_x,k_y)^\dagger = \bar{v}(k_x,k_y)\sigma_0$, and where $\bar{u}$ and $\bar{v}$ are the Hermitian complex conjugate of $u$ and $v$, respectively. One should notice that $\sigma_0$ represents a certain metric for the non-unitary representations. (An example of using a special metric for non-unitary representations is demonstrated in [16]). Following these definitions we get:

$$\bar{u}(k_x,k_y)\sigma_0 u(k_x,k_y) = (1,\bar{G}_1)\begin{pmatrix}1 & 0\\ 0 & -1\end{pmatrix}\begin{pmatrix}1\\ G_1\end{pmatrix} = 1 - |G_1|^2 \quad, \tag{31}$$

so that the normalized wavefunction (indicated by the subscript N) is given by

$$u_N = \begin{pmatrix}1\\ G_1\end{pmatrix} / \sqrt{1-|G_1|^2} \quad. \tag{32}$$

For orthogonality relation we obtain by using (15) and (23)

$$\bar{u}(k_x,k_y)\sigma_0 v(k_x,k_y) = (1,\bar{G}_1)\begin{pmatrix}1 & 0\\ 0 & -1\end{pmatrix}\begin{pmatrix}G_2\\ 1\end{pmatrix} = G_2 - \bar{G}_1 = 0 \quad. \tag{33}$$

For normalization of $v(k_x,k_y)$, representing the negative energy solutions, we get:

$$\bar{v}(k_x,k_y)\sigma_0 v(k_x,k_y) = (\bar{G}_2,1)\begin{pmatrix}1 & 0\\ 0 & -1\end{pmatrix}\begin{pmatrix}G_2\\ 1\end{pmatrix} = -1(1-|G_2|^2) \quad. \tag{34}$$

where $|G_2| = |G_1|$. The normalized $v_N(k_x,k_y)$ function is then given by

$$v_N(k_x,k_y) = \begin{pmatrix}G_2\\ 1\end{pmatrix} / \sqrt{1-|G_2|^2} \quad, \tag{35}$$

Summarizing the ortho-normalization relations we have:

$$u_N^\dagger(k_x,k_y) v_N(k_x,k_y) = v_N^\dagger(k_x,k_y) u_N(k_x,k_y) = 0 \quad, \tag{36}$$

$$u_N^\dagger(k_x,k_y) u_N(k_x,k_y) = 1 \quad, \quad v_N^\dagger(k_x,k_y) v_N(k_x,k_y) = -1 \quad. \tag{37}$$

The ortho-normalization relations of (36-37) will be used in the next Section where we develop the SO(2,1) relativistic quantum field theory in which $\hat{\Psi}$ and $\hat{\bar{\Psi}}$ are field operators, leading to propagating plane waves, which are described with the use of creation and annihilation operators.

## 3. Quantization of the SO(2,1) Relativistic Equation

In the "simple" approach presented in the previous Section we get the energy $\hbar\omega$ for the state $\Psi_+$, and the energy $-\hbar\omega$ for the state represented by $\Psi_-$, where $\omega$ is the positive value given by Eq. (11).

This simple approach of having 'holes' with negative energies raises a fundamental problem as the energy is not bounded from below, which cannot be correct. In analogy to the ordinary Dirac equation the solution for this problem is to use quantum field theory in which we expand the quantum field operators $\hat{\Psi}$ and its Hermitian conjugate $\hat{\bar{\Psi}}$ into linear combinations of plane waves solutions, using (9-10), and quantize the fields with creation and annihilation operators. In this theory creation operators $\hat{d}^\dagger$ for positrons are combined with annihilation operators $\hat{b}$ for electrons, and vice versa. The negative energy solutions for electrons will represent positive energy solutions for positrons so that the problem of having negative energy solutions is eliminated. In order to implement this theory one needs to assume that the creation and annihilation operators will satisfy anti-commutation relations, which have very important consequences for electrons and positron statistics [17,18]. The quantum field theory for the SO(2,1) relativistc equation is developed as follows.

The classical *free field Lagrangian* density of the three-dimensional Dirac equation is given by

$$L = -c\hbar \bar{\Psi} \gamma_\mu \frac{\partial}{\partial x_\mu} \Psi - mc^2 \bar{\Psi}\Psi \qquad . \tag{38}$$

The independent fields are considered to be the two components of $\Psi$ and the two components of $\bar{\Psi}$. The Euler-Lagrange equation using independent $\bar{\Psi}$ fields is given as

$$\frac{\partial}{\partial x_\mu}\left(\frac{\partial L}{\partial \bar{\Psi}/\partial x_\mu}\right) - \frac{\partial L}{\partial \bar{\Psi}} = 0 \qquad . \tag{39}$$

Since there is no derivative of $\bar{\Psi}$ in the Lagrangian we get the simple equation

$$\frac{\partial L}{\partial \bar{\Psi}} = 0 \qquad . \tag{40}$$

By substituting Eq. (38) into Eq. (40) we get:

$$-c\hbar \gamma_\mu \frac{\partial}{\partial x_\mu}\Psi - mc^2 \Psi = 0 \Rightarrow \left(\gamma_\mu \frac{\partial}{\partial x_\mu} + \frac{mc}{\hbar}\right)\Psi = 0 \quad, \tag{41}$$

which is the relativistic equation given in Eq. (3), indicating that the Lagrangian (38) is the right one. To compute the Hamiltonian density, we start by finding the momenta conjugate to the fields $\Psi$ given as

$$\Pi = \frac{\partial L}{\partial\left(\frac{\partial \Psi}{\partial t}\right)} = -c\hbar \bar{\Psi} \gamma_0 / c \qquad . \tag{42}$$

The Hamiltonian density is then given by

$$H = \Pi \frac{\partial \Psi}{\partial t} - L = -c\hbar \bar{\Psi} \gamma_0 \frac{1}{c} \frac{\partial \Psi}{\partial t} + c\hbar \bar{\Psi} \gamma_0 \frac{\partial}{\partial x_0} \Psi + c\hbar \bar{\Psi} \gamma_1 \frac{\partial \Psi}{\partial x} + c\hbar \bar{\Psi} \gamma_2 \frac{\partial \Psi}{\partial y} \quad , \tag{43}$$

where the first two terms on the right side of Eq. (43) are cancelled so that

$$H = c\hbar \left[ \bar{\Psi} \left( \gamma_1 \frac{\partial}{\partial x} + \gamma_2 \frac{\partial}{\partial y} + mc/\hbar \right) \Psi \right] \quad . \tag{44}$$

Using (1-3) we apply the transformation

$$\left( \gamma_1 \frac{\partial}{\partial x} + \gamma_2 \frac{\partial}{\partial y} + mc/\hbar \right) \Psi = \left( \frac{i\sigma_0}{c} \right) \frac{\partial}{\partial t} \Psi \quad , \tag{45}$$

so that (44) can be transformed into the simpler form

$$H = \hbar \left[ \bar{\Psi} (i\sigma_0) \frac{\partial}{\partial t} \Psi \right] \quad . \tag{46}$$

In analogy with the quantum field theory of the Dirac equation [6-8] in which $\Psi$ and $\bar{\Psi}$ become field operators we get here for the SO(2,1) relativistic equation::

$$\hat{\Psi}(x,y,t) = \iint dk_x dk_y \left( \frac{1}{2\pi} \right) \begin{bmatrix} \hat{b}(k_x,k_y) u_N(k_x,k_y) \exp[i(k_x x + k_y y - \omega t)] + \\ \hat{d}^\dagger(k_x,k_y) v_N(k_x,k_y) \exp[i(k_x x + k_y y - \omega t)] \end{bmatrix} \quad , \tag{47}$$

$$\bar{\hat{\Psi}}(x,y,t) = \iint dk_x dk_y \left( \frac{1}{2\pi} \right) \begin{bmatrix} \hat{b}^\dagger(k_x,k_y) \bar{u}_N(k_x,k_y) \exp[-i(k_x x + k_y y - \omega t)] + \\ \hat{d}(k_x,k_y) \bar{v}_N(k_x,k_y) \exp[-i(k_x x + k_y y - \omega t)] \end{bmatrix} \quad . \tag{48}$$

We take into account that all operators operating on the electrons commute with all operators operating on the holes. The form of $\hat{\Psi}(x,t)$ and $\bar{\hat{\Psi}}(x,t)$ is adapted here for the SO(2,1) Dirac equation. The creation and annihilation operators satisfy the anti-commutation relations:

$$\{\hat{b},\hat{b}\} = \{\hat{d},\hat{d}\} = \{\hat{b},\hat{d}\} = \{\hat{b}^\dagger,\hat{b}^\dagger\} = \{\hat{d}^\dagger,\hat{d}^\dagger\} = \{\hat{b}^\dagger,\hat{d}^\dagger\} = \{\hat{b},\hat{d}^\dagger\} = \{\hat{d},\hat{b}^\dagger\} = 0 \quad , \tag{49}$$

$$\{\hat{b}(k_x,k_y),\hat{b}^\dagger(k'_x,k'_y)\} = \{\hat{d}(k_x,k_y),\hat{d}^\dagger(k'_x,k'_y)\} = \delta(k_x - k'_x)\delta(k_y - k'_y) \quad . \tag{50}$$

In the present quantum field theory (46) represents Hamiltonian density and the Hamiltonian is obtained by integrating the Hamiltonian density over the $(x,y)$ plane so that

$$H = \iint dxdy \hbar \left[ \bar{\hat{\Psi}}(x,y,t) i\sigma_0 \frac{\partial}{\partial t} \hat{\Psi}(x,y,t) \right] = \iint dxdy \hbar \bar{\hat{\Psi}}(x,y,t) \sigma_0 \omega(k_x,k_y) \hat{\Psi}(x,y,t). \tag{51}$$

We notice that in (51) the metric $\sigma_0$ is appearing between $\overline{\hat{\Psi}}(x,y,t)$ and $\hat{\Psi}(x,y,t)$. We use here a special form for quantizing $\hat{\Psi}(x,y,t)$ and $\hat{\Psi}^\dagger(x,y,t)$ in the plane, given by (47-48), and where $u_N(k_x,k_y)$ and $v_N(k_x,k_y)$ satisfy the ortho-normalization conditions of (36-37), and where the anti-commutation relations of (49-50) are satisfied. We substitute $\hat{\Psi}(x,y,t)$ and $\overline{\hat{\Psi}}(x,y,t)$ from (47-48) into (51) and simplify further the expression for the Hamiltonian by performing first the integration over the x,y coordinates. Such integration leads to the delta function multiplication $\delta(k_x - k'_x)\delta(k_y - k'_y)$ so that by using these delta functions the integration in momentum space is reduced to $\iint dk_x dk_y$ and then we get:

$$H = \iint dk_x dk_y \hbar\omega(k_x,k_y)\left[\hat{b}^\dagger(k_x,k_y)\hat{b}(k_x,k_y) - \hat{d}(k_x,k_y)\hat{d}^\dagger(k_x,k_y)\right] . \tag{52}$$

The negative term in Eq. (52) follows from the ortho-normalization conditions of (34-37). Using the anti-commutation relations of (50) we can substitute in (52) the relation

$$\hat{d}(k_x,k_y)\hat{d}^\dagger(k_x,k_y) = -\hat{d}^\dagger(k_x,k_y)\hat{d}(k_x,k_y) + \delta(k_x)\delta(k_y) \quad , \tag{53}$$

and by neglecting the terms with zero momentum we get

$$H' = \iint dk_x dk_y \hbar\omega(k_x,k_y)\left[\hat{b}^\dagger(k_x,k_y)\hat{b}(k_x,k_y) + \hat{d}^\dagger(k_x,k_y)d(k_x,k_y)\right] . \tag{54}$$

$H'$ is obviously a non-negative operator. Assuming that $n^\pm(k_x,k_y)$ are the occupation numbers of electrons and positrons with momenta $(k_x,k_y)$ then the energy of the state is given by

$$E' = \iint dk_x dk_y \hbar\omega(k_x,k_y)\left[n^+(k_x,k_y) + n^-(k_x,k_y)\right] \geq 0 . \tag{55}$$

We should take into account that the canonical quantization rules lead to the following anti-commutation relations:

$$\{\Psi,\Psi\} = \{\Psi^\dagger,\Psi^\dagger\} = 0 \quad , \quad \{\Psi(x,y,t),\Psi^\dagger(x',y',t)\} = \delta(x-x')\delta(y-y') \tag{56}$$

where

$$\Psi^\dagger(x,y,t) = \overline{\Psi}(x,y,t)\sigma_0 . \tag{57}$$

In a similar way to the calculation made for the Hamiltonian, which are based on the ortho-normalization relations, we need also here to introduce the metric $\sigma_0$ so that the integration is made between $\Psi^\dagger(x',y',t) = \overline{\Psi}(x',y',t)\sigma_0$ and $\Psi(x,y,t)$.

The above theoretical analysis is important conceptually as, in analogy with the ordinary Dirac equation, it shows that the use of SO(2,1) Dirac equation becomes consistent only under the assumption of the above anti-commutation relations. The use of SO(2,1) Dirac equation is important for certain applications in solid state physics as will be demonstrated in the next Section.

## 4. Applications

### a) The non-relativistic limit of the SO(2,1) Dirac equation

We will derive the non-relativistic limit of the SO(2,1) Dirac equation by substituting in (1) the definitions:

$$\tilde{\Psi}_\pm = \exp\left[\left(-imc^2/\hbar\right)t\right]\Psi_\pm \quad , \tag{58}$$

and expand it into the following two equations:

$$-\frac{i}{c}\frac{\partial}{\partial t}\tilde{\Psi}_+ + \left(\frac{\partial}{\partial x} - i\frac{\partial}{\partial y}\right)\tilde{\Psi}_- = 0 \quad , \tag{59}$$

$$\frac{i}{c}\frac{\partial}{\partial t}\tilde{\Psi}_- + \left(\frac{\partial}{\partial x} + i\frac{\partial}{\partial y}\right)\tilde{\Psi}_+ + \left(\frac{2mc}{\hbar}\right)\tilde{\Psi}_- = 0 \quad . \tag{60}$$

Due to the large coefficient $\left(\frac{2mc}{\hbar}\right)$, in the non-relativistic limit, we can neglect the time derivative of the small component $\tilde{\Psi}_-$ and get from Eq. (60) the approximation:

$$\tilde{\Psi}_- = \left(\frac{\partial}{\partial x} + i\frac{\partial}{\partial y}\right)\tilde{\Psi}_+ \Big/ \left(\frac{2mc}{\hbar}\right) \quad . \tag{61}$$

Substituting (61) into (59) we get:

$$-\frac{i}{c}\frac{\partial}{\partial t}\tilde{\Psi}_+ + \left[\left(\frac{\partial}{\partial x} - i\frac{\partial}{\partial y}\right)\left(\frac{\partial}{\partial x} + i\frac{\partial}{\partial y}\right)\right]\tilde{\Psi}_+ \Big/ \left(\frac{2mc}{\hbar}\right) = 0 \quad . \tag{62}$$

Multiplying (62) by $c\hbar$ and expanding the terms in the square brackets we obtain

$$i\hbar\frac{\partial}{\partial t}\tilde{\Psi}_+ = -\left[\frac{\hbar^2}{2m}\left(\frac{\partial^2}{\partial x^2} + \frac{\partial^2}{\partial y^2}\right)\right]\tilde{\Psi}_+ \quad , \tag{63}$$

which is the Schrodinger equation with the kinetic terms in the $(x, y)$ plane.

In order to study the interaction of the present non-relativistic particle with an external classical electro-magnetic field, characterized by a potential $A_\mu(x, y), (\mu = 0, 1, 2)$, we use the *minimal coupling* prescription [8] by which:

$$\frac{\partial}{\partial x} \Rightarrow \frac{\partial}{\partial x} + ieA_x \quad , \quad \frac{\partial}{\partial y} \Rightarrow \frac{\partial}{\partial y} + ieA_y \quad , \quad \frac{1}{c}\frac{\partial}{\partial t} \Rightarrow \frac{1}{c}\frac{\partial}{\partial t} + ieA_0 \quad , \tag{64}$$

where $e$ is the particle charge. Then (63) is transformed into

$$i\hbar \frac{\partial}{\partial t}\tilde{\Psi}_+ = -\left(\frac{\hbar^2}{2m}\right)\left[\left(\frac{\partial}{\partial x} + ieA_x\right)^2 + \left(\frac{\partial}{\partial y} + ieA_y\right)^2 + eA_0\right]\tilde{\Psi}_+ \quad . \tag{65}$$

The interesting feature which we have obtained here is that the present SO(2,1) Dirac equation is based on two-dimensional wavefunctions and that the non-relativistic limit of this equation leads to (65), which is the basis equation for treating quantum Hall effects in the plane [10-12]. So, we find that the present model is consistent with other derivations of (65), which are starting already from the beginning with Schrodinger equation.

**b) A Relativistic Effect Related to Holes-Electrons pairs**

We notice that the creation and annihilation operators applied in (54) satisfy the anti-commutation relations which have important implications for particle statistics [17,18]. A possible representation for hole-particle production and annihilation perturbation is given as:

$$\hat{O}_{hole-elec.} = \hat{b}^\dagger(k_x, k_y)\hat{d}^\dagger(-k_x, -k_y) + \hat{b}(k_x, k_y)\hat{d}(-k_x, -k_y) \quad . \tag{66}$$

One should notice the following characteristics of the operator $\hat{O}_{hole-elec.}$:

a) In spite of the non-unitary representations which have used for the SO(2,1) Dirac equation both the Hamiltonian of (54) and the operator $\hat{O}_{hole-elec.}$ are Hermitian.

b) Both Hamiltonian $H'$ and $\hat{O}_{hole-elec.}$ are described in momentum space so that they describe non-local effects.

c) Both Hamiltonian $H'$ and $\hat{O}_{hole-elec.}$ preserve the charge and momentum of the system.

d) While the creation and annihilation $\hat{b}$ and $\hat{d}$ operators satisfy the anti-commutation relations given by (47-48) the operator $\hat{O}_{hole-elec.}$ satisfy the bosonic commutation-relations:

$$\left[\hat{b}(k_x, k_y)\hat{d}(-k_x, -k_y), \hat{b}^\dagger(k'_x, k'_y)\hat{d}^\dagger(-k'_x, -k'_y)\right] = \delta(k_x - k'_x)\delta(k_y - k'_y) \quad . \tag{67}$$

In deriving (67) we have used the anti-commutation relations of (50), twice. One should notice that the anti-commutation relations are changing the sign in (65) twice so that by multiplication of the anti-commutation relations we get bosonic relations. The coupling mechanism of (66) between two particles in the plane, by the use of the SO(2,1) Dirac equation, is different from coupling mechanism in the

x,y,z,t space, as in the present case it includes two-dimensional wavefunctions. We find also that the operator representing the number of hole-electron pairs in such system is given as

$$\hat{N}_{hole-elec} = \hat{b}^{\dagger}(k_x,k_y)\hat{d}^{\dagger}(-k_x,-k_y)\hat{b}(k_x,k_y)\hat{d}(-k_x,-k_y) \quad . \tag{68}$$

This operator represents the number of hole-electron pairs which couples the momenta $(k_x,k_y)$ of the electrons with that of $(-k_x,-k_y)$ of the holes. By integrating this operator over all possible hole-electron coupling in momentum space, we represent the number operator for the total hole-electron pairs. The interesting result in the above formalism is that the operator $\hat{N}_{hole-elec}$ is a bosonic operator, due to the commutation-relations (67) and there is no restriction on this number as Pauli exclusion principle does not apply here. While we have concentrated here on hole-electron pair interaction [13], such effect is an essentiall feature of various two-particles interactions which can lead to bosonic states, although each particle operator by itself satisfy the anti-commutation relations.

## 5. Summary and conclusions

It has been shown that the Dirac equation in x,y,z,t space, which is based on the SO(3,1) group, can be varied to a Dirac equation in x,y,t space, which is based on the SO(2,1) group. Thus we are using reduced Dirac equation of the form of Eq. (1), in the low dimensional space, where the coordinate z is eliminated, the Pauli matrices are exchanging the $\gamma$ matrices and the present wavefunctions are two-dimensional. The general un-normalized wavefunctions solutions for the varied Dirac equation, are given as the wavefunction $\Psi_+$ of Eq. (16) with the parameter $G_1(k_x,k_y)$ calculated in Eq. (22) ,and the wavefunction $\Psi_-$ of Eq. (24) with the parameter $G_2(k_x,k_y)$ calculated in Eq. (30). Since we use finite non-unitary representations, in analogous way to that of the Dirac equation, we insert a certain metric by which the ortho-normalization relations are calculated and given in Eqs. (36-37). We get positive and negative energy solutions in analogous way to the ordinary Dirac equation.

In analogy to the ordinary Dirac equation we use quantum field theory in Section 3 by which the quantum wavefunctions are operators and expand them into linear combinations of plane wave solutions with creation and annihilation operators satisfying the anti-commutation relations given by (49-50). In this theory the creation operators $d^{\dagger}$ for positrons are combined with annihilation operators $\hat{b}$ for electrons, and vice versa, so that negative energy solutions for electrons represent positive energy solutions for positrons. These solutions are based on two-dimensional wavefunctions.

It has been shown in Section 4 that the relativistic limit of the SO(2,1) Dirac equation leads to the basic equations of quantum Hall effects, in the plane, which are consistent with well known results in this field. It has been shown by using a quantum field theory, that while one-electron operators satisfy the anti-commutation relations, two-particle operators can lead to bosonic interactions, (see e.g.[13]), including the bosonic commutation relations.

## 5. References

1. Leinhaas, J.M. Myrheim, J. : Nuvo Cimento **37** 1 (1977).

2. Myrheim,J.: Anyons. In: Topological aspects of low dimensional systems (Comtet, A., Jolicoeur,T., Quary,S., David,F., (eds), Springer , Berlin, pp. 268-413 (1998).

3. Brennen,G.V.,Pachos,J.K.: Proc.R.Soc. A , 464 (2008) .

4. Nayak, C., Simon, S.H., Stern A., Freedman, M., Sarma, S.D.: Rev.Mod.Phys**. 80**, 1083 (2008).

5. Ben-Aryeh, Y. : J.Opt.B:Quantum Semiclass. Opt. **6**, R1-R19 (2004).

6. Chang, S.J.: Introduction to quantum field theory. World Scientific. Singapore (1990).

7. Kaku, M.: Quantum field theory. Oxford University Press . New York ( 1993).

8. Itzykson, C., Zuber, J-B : Quantum field theory. McGraw Hill. New York (1980).

9. Kim, Y.S., Noz, M.E. : Theory and Applications of the Poincare Group. Kluwer. Dordrecht (1986); Kim, Y.S., Noz, M.E. : Phase Space Picture of Quantum Mechanics.World Scientific. Singapore (1991).

10. Yoshioka, D.: The quantum Hall effect . Springer. Berlin (2002).

11. Chackraborty ,T., Pietilainen, P.: The Quantum Hall effects .Springer, Berlin (1995).

12. Prange, R.E., Girvin S.M. (eds.): The quantum Hall effect . Springer. Berlin (1995).

13. Brinkman, A., Hilgenkamp, H.: Electron-hole coupling in high TC materials : Arxiv: cond-mat/0503368 [con- mat.supr-cond] (2005.)

14. Wigner, E., Annals of Mathematics **40**, 149 (1939).

15. Bargmann, W.: Annals of Mathematics **48**, 568 (1947).

16. Ben-Aryeh, Y., Mann A., Yaakov, I.: J.Phys.A: Math.Gen. **37**, 12059 (2004).

17. Duck, I., Sudarshan, E.C.G.: Pauli and the spin-statistics theorem .World Scientific. Singapore (1995).

18. Streater, R.F., Wightman, A.S. : PCT, spin and statistics and all that. Princeton University Press. Princeton (1980).